\documentclass[%
 reprint,
superscriptaddress,
preprintnumbers,
nofootinbib,
 amsmath,amssymb,
]{revtex4-2}
\usepackage{cancel}
\usepackage{dblfloatfix}
\usepackage{xcolor}
\usepackage{booktabs}
\usepackage{graphicx}
\usepackage{bm}
\usepackage{hyperref}

\newcommand{\planck}[0]{\textit{Planck}}
\newcommand{\wmap}[0]{\textit{WMAP}}
\newcommand{\bini}{\beta_{\rm ini}}
\newcommand{\brec}{b_{\rm rec}}
\newcommand{\mphi}{m_\phi}

\begin{document}

\title{Constraints on the mass of an axion-like field from parity-violating signal in the ACT DR6 polarization spectra}

\author{Debabrata Adak}
\email{adak@iac.es}
\affiliation{Instituto de Astrof\'isica de Canarias, E-38205 La Laguna, Tenerife, Spain}
\affiliation{Departamento de Astrof\'{\i}sica, Universidad de La Laguna, E-38206 La Laguna, Tenerife, Spain}

\date{\today}

\begin{abstract}
The polarization data of the Atacama Cosmology Telescope (ACT) Data Release 6 (DR6) show a hint of isotropic cosmic birefringence with a rotation angle $\beta\simeq0.2^\circ$. If the rotation is caused by an axion-like field coupled to photons, the angle can change during recombination, depending on how the field evolves. The shape of the parity-odd $EB$ and $TB$ power spectra then depends on the field mass $\mphi$, which the data can therefore constrain.
We compute the lensed $EB$ and $TB$ spectra for a quadratic potential with a modified Boltzmann code that follows the time-dependent rotation along the line of sight, for 216 masses in $-30\le\log_{10}(\mphi/{\rm eV})\le-26.5$. Using these spectra as templates in the Minami--Komatsu estimator, we fit the ACT DR6 cross-array-band $EB$ and $TB$ power spectra for the initial rotation angle together with polarization miscalibration angles constrained by priors from the ACT optics model. In the light-mass limit we recover the constant-angle result, $\beta=0.204^\circ\pm0.058^\circ$. Masses $\log_{10}(\mphi/{\rm eV})\lesssim-28.5$ cannot be distinguished from a constant rotation. For larger masses the minimum $\chi^2$ oscillates with the phase of the field at recombination, and masses for which the field crosses zero during recombination are disfavored by $\Delta\chi^2\simeq5$--$6$. In a Bayesian analysis, the marginal posterior excludes $\log_{10}(\mphi/{\rm eV})\in[-27.96,-27.74]$, $[-27.59,-27.45]$, $[-27.38,-27.27]$, and $[-27.24,-26.5]$ at 95\% confidence. Marginalizing over the $\Lambda$CDM parameters does not change these intervals. The three narrow windows between them, centered at $\log_{10}(\mphi/{\rm eV})\simeq-27.70$, $-27.42$, and $-27.26$, fit the data as well as a constant rotation. This pattern agrees with that found from \planck\ data.
\end{abstract}

\maketitle

\section{\label{sec:intro}Introduction}

A rotation of the plane of linear polarization of the cosmic microwave background (CMB), known as cosmic birefringence, would be a signature of parity-violating physics beyond the standard model~\cite{Carroll1990,Carroll1998,Komatsu2022review}.
An isotropic rotation by an angle $\beta$ rotates the plane of CMB polarization and generates odd-parity signals, $EB$ and $TB$ correlations.
Because a miscalibration of the polarization angles of the detectors produces the same signal, a measurement of $\beta$ requires independent information on the miscalibration, either from the polarized Galactic foreground emission~\cite{MinamiKomatsu2020a,MinamiKomatsu2020b} or from a model of the instrument \cite{DPK2025}. Alternatively, an internal linear combination of the $E$- and $B$-mode maps can separate the frequency-dependent miscalibration from the frequency-independent birefringence~\cite{Remazeilles:2025}.
Analyses of the \planck\ data yield $\beta=0.35^\circ\pm0.14^\circ$~\cite{MinamiKomatsu2020b} and, in combination with \wmap, $\beta=0.34^\circ\pm0.09^\circ$~\cite{EskiltKomatsu2022}.
Using the cross-array-band $EB$ and $TB$ power spectra of the Atacama Cosmology Telescope (ACT) Data Release 6 (DR6)~\cite{ACTDR6Louis} and priors on the miscalibration angles derived from the ACT optics model, Ref.~\cite{DPK2025} found $\beta=0.215^\circ\pm0.074^\circ$. A joint analysis of the \planck\ and ACT DR6 polarization maps gives $\beta=0.277^\circ\pm0.057^\circ$~\cite{Eskilt2026}. The rotation angle can also vary across the sky, which is called anisotropic birefringence~\cite{Pospelov2009,Kamionkowski2009,Caldwell2011}. All measurements so far are consistent with no anisotropic rotation~\cite{POLARBEAR2015,BK2017,BK2023,Namikawa2020,Bianchini2020,Gruppuso2020,Bortolami2022,Lonappan2025,Balkenhol2025,Kramer2026,Coerver2026}.

Radio galaxies provide an independent test of birefringence with different systematics, because the intrinsic polarization angle of a radio galaxy is statistically aligned with its position angle on the sky. Using 16,649 radio sources at $z\sim0.5$ from the Polarization Sky Survey of the Universe’s Magnetism (POSSUM), Ref.~\cite{Bruggen2026} found $\beta=0.48^\circ\pm0.35^\circ$, in agreement with the CMB measurements in sign and amplitude.

A natural source of cosmic birefringence is a pseudoscalar field $\phi$ coupled to photons through Chern-Simons interaction, $\mathcal{L}\supset-\tfrac14 g\phi F_{\mu\nu}\tilde F^{\mu\nu}$, which rotates the polarization of a photon emitted at conformal time $\eta$ by $\beta(\eta)=(g/2)[\phi(\eta_0)-\phi(\eta)]$, where $\eta_0$ is the present time \cite{Nakai:2023}. Proposed candidates for this field include dark energy~\cite{Fujita2021}, early dark energy~\cite{Capparelli2020}, axion-like dark matter, and domain walls of axion-like fields~\cite{TakahashiYin2021,Lee2025}.
The rotation of the CMB polarization depends on when and how the field evolves.
A field that is still frozen at recombination rotates all CMB photons by the same angle, whereas a field whose mass is comparable to the Hubble rate at recombination changes the angle while the CMB is emitted and gives the $EB$ spectrum a different shape~\cite{Nakatsuka2022,Murai2023,Galaverni:2023}.
Ref.~\cite{NMN2025} used this effect to constrain the mass of an axion-like field with the \planck\ data at $\ell < 1500$ and excluded several masses at more than $2\sigma$.

In this paper, we carry out the corresponding analysis with the ACT DR6 polarization data at $600\lesssim\ell\lesssim7900$, which complements the \planck-based analysis of Ref.~\cite{NMN2025}. We use the pipeline of Ref.~\cite{DPK2025}. Axion-like fields with masses of $10^{-32}$--$10^{-20}$\,eV have been constrained in many previous works with ACT and other data through their abundance as dark matter~\cite{ACTDR6Extended,Lague2026,Gaughan2026,Verdiani2026} or a rotation that oscillates during the observations~\cite{Huber2026}. The ACT $EB$ spectra have also been used to test early dark energy models with a photon coupling~\cite{Yin2025,ZhangYin2026}. In this paper, we constrain the mass of the axion-like field responsible for the observed rotation from the shape of the ACT $EB$ and $TB$ power spectra.

The paper is organized as follows. Section~\ref{sec:model} describes the model and the computation of the templates using modified \texttt{CAMB} and our choice of axion-mass range analyzed in this paper. In Sec.~\ref{sec:data} we describe the data, parameter priors, and the likelihood. Section~\ref{sec:validation} validates the pipeline. In  Sec.~\ref{sec:results} we present the results, and in Sec.~\ref{sec:discussion} we summarize results and conclude.
Appendix~\ref{app:posterior} derives the marginal posterior of the mass, Appendix~\ref{app:stacked} shows the stacked power spectra, and Appendix~\ref{app:lcdm} describes the marginalization over the cosmological parameters.

\section{\label{sec:model}Model and templates}

\subsection{\label{sec:rotation}Rotation angle for an evolving field}
The coupling $-\tfrac14 g\phi F_{\mu\nu}\tilde F^{\mu\nu}$ gives the two helicity states of a photon  $\omega_\pm$ \citep{Carroll1990}.
The accumulated phase difference rotates the plane of linear polarization of a photon emitted at conformal time $\eta$ and observed today $\eta_0$, by~\cite{Carroll1990,Nakatsuka2022}
\begin{equation}
\beta(\eta)=-\int_\eta^{\eta_0}d\eta_1\,\frac{\omega_+-\omega_-}{2}=\frac{g}{2}\left[\phi(\eta_0)-\phi(\eta)\right],
\label{eq:beta_eta}
\end{equation}
where $\beta>0$ corresponds to a clockwise rotation of the linear polarisation.
We consider a field with the potential $V(\phi)=\mphi^2\phi^2/2$, evolved according to equation of motion~\cite{Nakatsuka2022} in conformal time $\eta$ as
\begin{equation}
\phi''+2\frac{a'}{a}\phi'+a^2\mphi^2\phi=0,
\label{eq:eom}
\end{equation}
where primes denote derivatives with respect to $\eta$.
With $dt=a\,d\eta$ and $H=a'/a^2$, Eq.~\eqref{eq:eom} is equivalent to $\ddot\phi+3H\dot\phi+\mphi^2\phi=0$ in cosmic time.
We neglect the term $a^2g\phi F_{\mu\nu}\tilde F^{\mu\nu}$ in Eq.~\eqref{eq:eom}, which is quadratic in the cosmological perturbations. This is allowed if the energy density of the field is a negligible fraction of the total, which still allows a sizable rotation~\cite{Nakatsuka2022}.
We consider field starts at rest, $\phi=\phi_{\rm ini}$ and $\phi'=0$, at a time when $H\gg\mphi$.
\begin{figure}[t]
\includegraphics[width=\linewidth]{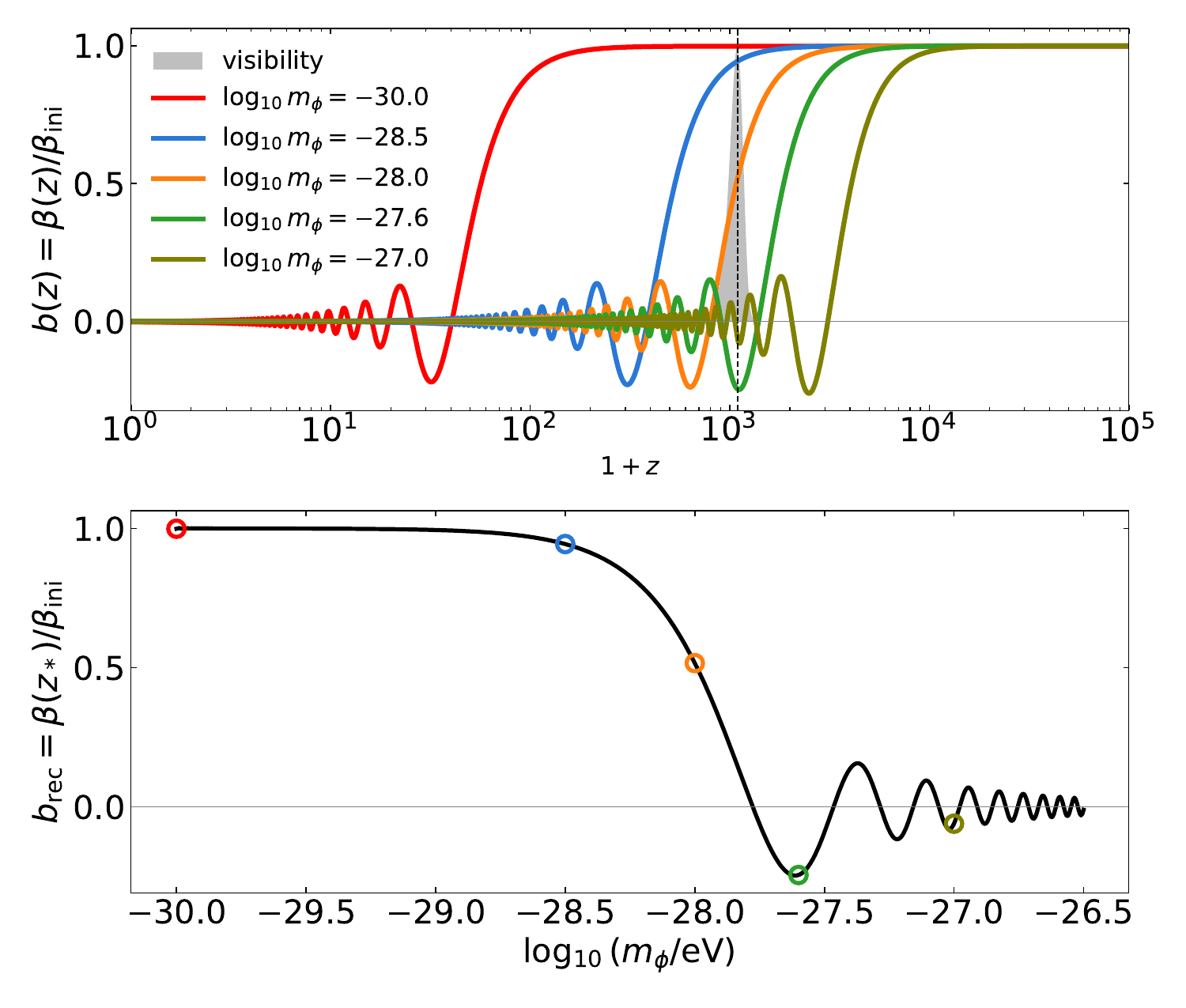}
\caption{Rotation history of the field for $\Lambda$CDM background.
Top: $b(z)=\beta(z)/\bini$, fraction of the initial rotation that remains at redshift $z$,  for four different masses.
The gray band shows the visibility function, normalized to its maximum value, and the dashed line marks $z_*=1090$.
Light fields are frozen through recombination ($b\simeq1$), whereas heavier fields start to roll before or during recombination and then oscillate with a decaying amplitude. Bottom: $\brec=b(z_*)$ as a function of mass. The circles highlight the masses used in the top panel.}
\label{fig:brec} 
\end{figure}
Writing $\phi=\phi_{\rm ini}f$ and defining $\bini\equiv-g\phi_{\rm ini}/2$, Eq.~\eqref{eq:beta_eta} becomes~\cite{NMN2025}
\begin{equation}
\beta(z)=\bini\,[f(z)-f(0)]\equiv\bini\,b(z).
\label{eq:beta}
\end{equation}
The function $f$, and therefore $b$, depends only on the mass, with the initial condition $f=1$, $f'=0$.
For a given mass, the rotation history is thus fixed up to the amplitude $\bini$.
We define $\brec(\mphi)\equiv b(z_*)$ with $z_*=1090$, the fraction of the initial rotation that happens at recombination for a given mass $\mphi$.
The field is frozen while $\mphi\ll H$ and starts to roll when $\mphi\sim H$; afterward it oscillates with decaying amplitude. The top panel of Figure.~\ref{fig:brec} shows the evolution of $b(z)$ for several masses, and the bottom panel of Figure.~\ref{fig:brec} shows $\brec=b(z_*)$ as a function of mass. For $\log_{10}(\mphi/{\rm eV})\lesssim-28.5$ the field is still frozen at recombination and $\brec\simeq1$, so that the rotation is constant across the last-scattering surface.
Heavier fields start to roll before recombination, and $\brec$ oscillates around zero with a decaying amplitude.

We solve for $f$ in time grid of $x\equiv\ln a$,
\begin{equation}
\frac{d^2f}{dx^2}+\left(3+\frac{d\ln H}{d\ln a}\right)\frac{df}{dx}+\left(\frac{\mphi}{H}\right)^2f=0,
\label{eq:eom_lna}
\end{equation}
which is Eq.~\eqref{eq:eom} with $d/dt=H\,d/dx$, using $H(z)$ computed by \texttt{CAMB}~\cite{CAMB} for the same cosmology as the spectra.
The integration starts at $a=10^{-8}$ with $f=1$ and $df/dx=0$, and uses an eighth-order Runge-Kutta method with a relative tolerance of $10^{-10}$.
Once $\mphi/H$ exceeds $10^3$, the field oscillates many times per Hubble time. These fast oscillations average out in the line-of-sight integral. We therefore stop the integration there and set $f=0$ afterward.

\subsection{\label{sec:camb}Modifications to \texttt{CAMB}}
We implement the time-dependent rotation in \texttt{CAMBER}\footnote{\url{https://github.com/Debabrataadak/CAMBER.git}} (\texttt{CAMB} with \texttt{E}volving \texttt{R}otation), a modified version of the Boltzmann code \texttt{CAMB}~\cite{CAMB}, which takes $\beta(z)$ as a table and rotates the polarization at each time along the line of sight. 
Our implementation follows Ref.~\cite{Galaverni:2023}, who did a similar modification in \texttt{CAMB}  and Refs.\cite{Nakatsuka2022, Naokawa2023}, who did so in \texttt{CLASS} (\texttt{birefCLASS}) and added the lensing effect to the parity-odd spectra.

The coupling adds a term $\mp i g\phi'\,\Delta_{Q\pm iU}$ to the Boltzmann equation of the polarization \cite{Galaverni:2023,Naokawa2023,Liu2006}.
Integrating it along the line of sight  rotates the polarization generated at time $\eta$ by the angle $\beta(\eta)$ that the photon accumulates until today $\eta_0$,
\begin{equation}
\Delta^{E}_\ell(k)\pm i\Delta^{B}_\ell(k)=\int_0^{\eta_0}d\eta\,S_E(k,\eta)\,j_\ell[k(\eta_0-\eta)]\,e^{\pm 2i\beta(\eta)},
\label{eq:DeltaEB}
\end{equation}
where the $E$-mode source $S_E$ feeds both $E$ and $B$ modes. The real and imaginary parts are polarization scalar perturbations for $E$ and $B$ modes, with  $j_\ell$ indicating the spherical Bessel function of order $\ell$~\cite{Seljak1996}.
In \texttt{CAMBER}, we evaluate $\beta$ at each time step of the line-of-sight integration and compute the power spectra $C^{XY}_\ell$  following
\begin{equation}
C^{XY}_\ell=4\pi\int d\ln k\,\mathcal P(k)\,\Delta^X_\ell(k)\Delta^Y_\ell(k),
\label{eq:cl}
\end{equation}
where $X,Y\in\{T,E,B\}$. \texttt{CAMB} computes the lensed spectra with the curved-sky correlation-function method~\cite{Challinor2005}. We extended it to the parity-odd spectra following Ref.~\cite{Naokawa2023}. The \texttt{CAMBER} code returns lensed $TT, EE, BB, TE, EB $ and $TB$ power spectra.

For small angles, the spectra are linear in $\bini$.
Expanding Eq.~\eqref{eq:cl} with $\sin2\beta=2\beta-\tfrac43\beta^3+\ldots$ and $\cos2\beta=1-2\beta^2+\ldots$, and inserting Eq.~\eqref{eq:beta}, gives
\begin{equation}
C^{EB}_\ell=2\bini\,K^{EB}_\ell(\mphi)\left[1+\mathcal O(\beta^2)\right],
\label{eq:linear}
\end{equation}
with
\begin{equation}
K^{EB}_\ell=4\pi\int d\ln k\,\mathcal P(k)\left[\int d\eta\,S_E\,j_\ell\right]\left[\int d\eta\,S_E\,j_\ell\,b(\eta)\right],
\end{equation}
and analogously for $TB$ with the temperature transfer function in the first bracket.
Since $K_\ell$ depends only on the mass, the spectra per unit $\bini$ do not depend on the value of $\bini$ used to compute them.
For each mass, we run \texttt{CAMBER} with $\bini=0.1^\circ$ and define the templates
\begin{equation}
\mathcal{T}^{XB}_\ell(\mphi)\equiv\frac{D_\ell^{XB}(\mphi;\bini)}{\bini},\qquad X\in\{E,T\},
\label{eq:templates}
\end{equation}
where $D_\ell\equiv\ell(\ell+1)C_\ell/2\pi$ and $\bini$ is in radian, so that $\mathcal T_\ell$s are the lensed spectra per radian of $\bini$.  Figure~\ref{fig:templates} shows the templates of $\mathcal{T}^{EB}_\ell$ (top-panel) and $\mathcal{T}^{TB}_\ell$ (bottom-panel) per degree $\bini$ for representative four masses. As the mass increases, the amplitude is suppressed, the acoustic peaks shift, and the sign can change.

\subsection{\label{sec:grid}Cosmology, mass grid, and accuracy}

\begin{figure}
\centering
\includegraphics[width=\linewidth]{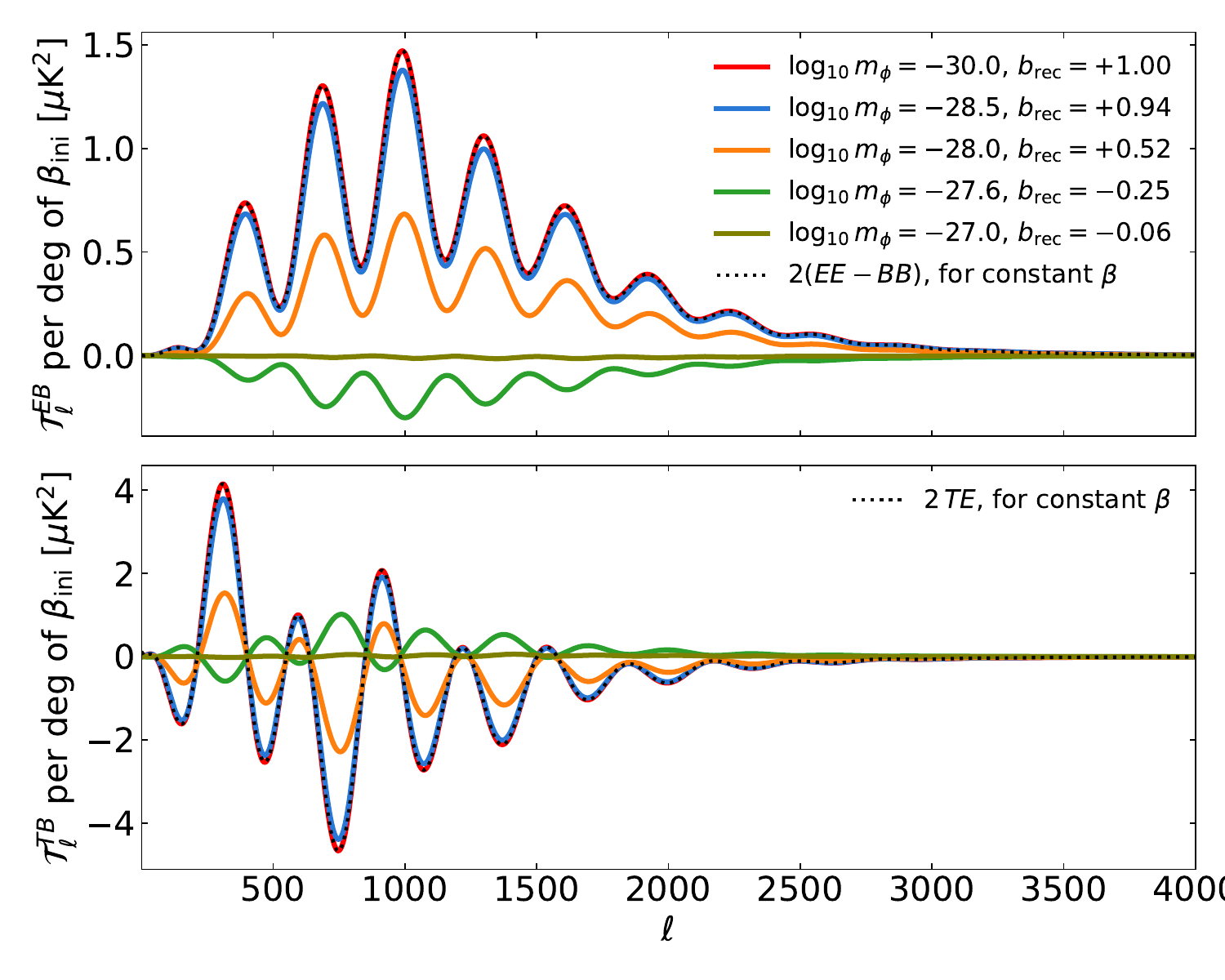}
\caption{\label{fig:templates} Lensed $EB$ (top) and $TB$ (bottom) templates per degree of $\bini$ for readability for several masses. The dotted lines show the corresponding templates per degree for a constant angle rotation, $2(D^{EE}_\ell-D^{BB}_\ell)$ (for $EB$) and $2D^{TE}_\ell$ (for $TB$). These spectra are computed from \texttt{CAMBER} code for ACT DR6 $\Lambda$CDM best-fit parameters.}
\end{figure}

The background cosmology is a flat $\Lambda$CDM model for the best-fit  parameters estimated fitting the ACT DR6 data \citep{Calabrese:2025}, 
We use 216 masses with steps of $0.1$, $0.02$, $0.01$, and $0.005$ in $\log_{10}(\mphi/{\rm eV})$ over the intervals $[-30,-28.5]$, $[-28.5,-27.5]$, $[-27.5,-27]$, and $[-27,-26.5]$, respectively.
These steps are chosen such that the phase of the field at recombination, which is approximately $\mphi/2H(z_*)$, changes by less than about $0.3$\,rad between neighboring masses. The spectra are computed up to $\ell=9000$ with the accuracy parameters \texttt{AccuracyBoost}$\,=2$ and \texttt{lSampleBoost}$\,=4$, with \texttt{lens\_potential\_accuracy}$\,=8$.
Figure.~\ref{fig:templates} shows the templates for representative masses. As the mass increases, the amplitude is suppressed, the acoustic peaks shift, and the sign can change.

\section{\label{sec:data}Data and likelihood}
\subsection{\label{sec:data_actrd6} DATA}
Our analysis uses the cross-array-band power spectra and covariance matrix of the ACT DR6\footnote{\url{https://lambda.gsfc.nasa.gov/product/act/act_dr6.02/}}~\cite{ACTDR6Louis} data.
Below we describe the relevant aspects for our analysis.

The ACT DR6 maps combined the observations made from 2017 to 2022 over $19\,000$\,deg$^2$, with a median depth of $10\,\mu{\rm K}$-arcmin and a resolution of about 1\,arcmin. Three polarization arrays (PAs) contribute. PA5 and PA6 each have an f090 (77--112\,GHz) and an f150 (124--172\,GHz) band, and PA4 has an f150 and an f220 (182--277\,GHz) band. We drop PA4 f150, which does not pass several null tests. PA4 f220 is not part of the ACT cosmological analysis, where it adds little constraining power, but we keep it, following Ref.~\cite{DPK2025}. In  total, five array-bands remain in our analysis, PA4 f220, PA5 f090, PA5 f150, PA6 f090, and PA6 f150, and each has its own miscalibration angle $\alpha_i$ \cite{Murphy:2024}.

The spectra are measured on the $10\,000$\,deg$^2$ left after masking the brightest Galactic regions and the extragalactic point sources. We use the binning and multipole cuts of Ref.~\cite{DPK2025}, listed in Table~\ref{tab:bins}, which give 1990 data points. As in Ref.~\cite{DPK2025}, we neglect the $EB$ and $TB$ correlations of Galactic dust, since at $\ell\ge600$ the polarization spectra are dominated by the CMB and the noise \cite{ACTDR6Louis}.

\begin{table}
\caption{\label{tab:bins} Centers of the first and last bins used, $b_{\rm min}$ and $b_{\rm max}$, and number of bins, $N_b$, for each array-band. The selection follows Ref.~\cite{DPK2025}.}
\begin{ruledtabular}
\begin{tabular}{lccc}
Array-band & $b_{\rm min}$ & $b_{\rm max}$ & $N_b$ \\
\hline
PA4 f220 & 1000.5 & 7525.5 & 39 \\
PA5 f090 & 1000.5 & 7525.5 & 39 \\
PA5 f150 & \phantom{0}800.5 & 7525.5 & 43 \\
PA6 f090 & 1000.5 & 7525.5 & 39 \\
PA6 f150 & \phantom{0}600.5 & 7525.5 & 47 \\
\end{tabular}
\end{ruledtabular}
\end{table}

\subsection{\label{sec:priors}Miscalibration priors}

Because an isotropic rotation and a miscalibration of the polarization angles affect $EB$ and $TB$ in the same way, we constrain the $\alpha_i$ with Gaussian priors of zero mean and standard deviations derived from the model of the ACT optics \cite{Murphy:2024}.
This model accounts for possible displacements and tilts of the lenses and for the properties of the antireflection coatings by means of polarization-sensitive ray tracing.
The resulting standard deviations are $\sigma_\alpha=0.09^\circ$ for PA4 and PA5 and $0.11^\circ$ for PA6, and the polarization angles vary by less than $0.01^\circ$ within each passband \cite{Murphy:2024}.
Ref.~\cite{DPK2025} considered independent angles ($\Pi_{\rm ind}$), a 90\% correlation between the bands of the same array ($\Pi_{\rm PA}$), which is their baseline because such bands share the same optics tube, and a 90\% correlation between the bands at the same frequency ($\Pi_\nu$).
They found $\beta=0.204^\circ\pm0.058^\circ$, $0.212^\circ\pm0.072^\circ$, and $0.212^\circ\pm0.072^\circ$, respectively, for three choices of the priors.
In this work, we adopt $\Pi_{\rm ind}$.

\subsection{\label{sec:likelihood}Likelihood}

We use the estimator of Refs.~\cite{MinamiKomatsu2020a,DPK2025}, with the posterior
\begin{equation}
-2\ln\mathcal P =\sum_{b,b'}\bm v_b^{\sf T}\,\mathbf M^{-1}_{bb'}\,\bm v_{b'}+\ln|\mathbf M|+\bm\alpha^{\sf T}\bm\Pi^{-1}\bm\alpha,
\label{eq:posterior}
\end{equation}
where the sum runs over the bins of Table~\ref{tab:bins} common to each pair of spectra,  $\mathbf M_{bb'}$ is their covariance, and $\bm\Pi$ is the prior covariance of $\bm\alpha\equiv(\alpha_i)$. For small angle approximation, $\bm v_b$ for each pair of  array-bands $(i,j)$ in bin $b$ contains
\begin{align}
v^{E_iB_j}_b&=D^{E_iB_j,{\rm o}}_b-2\alpha_jD^{E_iE_j,{\rm o}}_b\nonumber\\
&\quad+2\alpha_iD^{B_iB_j,{\rm o}}_b-A\,\mathcal T^{EB}_b(\mphi),\label{eq:vEB}\\
v^{T_iB_j}_b&=D^{T_iB_j,{\rm o}}_b-2\alpha_jD^{T_iE_j,{\rm o}}_b-A\,\mathcal T^{TB}_b(\mphi),\label{eq:vTB}
\end{align}
where the superscript o denotes the observed band powers and $\mathcal T_b$s are the templates of Eq.~\eqref{eq:templates} averaged over each bin.
In the constant-angle analysis of Ref.~\cite{DPK2025}, the last terms are $2\beta(D^{EE}_b-D^{BB}_b)$ and $2\beta D^{TE}_b$, computed from the $\Lambda$CDM best-fit. We replace them with $A\,\mathcal T^{EB}_b$ and $A\,\mathcal T^{TB}_b$ respectively, which is the only change we make to their  pipeline\footnote{\url{https://github.com/pdp79/act_dr6_analysis}}.
Because the miscalibration terms use the observed spectra, the CMB model enters only through the templates, and since templates are defined per radian of $\bini$, the amplitude is $A=\bini$ obtained by fitting the data.

The covariance $\mathbf M_{bb'}=\mathbf R\,{\rm Cov}(\bm D^{\rm o}_b,\bm D^{\rm o\,{\sf T}}_{b'})\,\mathbf R^{\sf T}$ where
\begin{equation}
\begin{split}
\mathbf R={\rm diag}\bigg[&\left(-\frac{\sin4\alpha_j}{\cos4\alpha_i+\cos4\alpha_j},\,\frac{\sin4\alpha_i}{\cos4\alpha_i+\cos4\alpha_j},\,1\right),\\
&\left(-\tan2\alpha_j,\,1\right)\bigg]
\end{split}
\end{equation}
with ${\rm Cov}(\bm D^{\rm o}_b,\bm D^{\rm o\,{\sf T}}_{b'})$ is taken from the published covariance of the observed $EE$, $BB$, $EB$, $TE$, and $TB$ spectra. 

At fixed $\mathbf M$ the residuals in  Eqs.~\eqref{eq:vEB} and \eqref{eq:vTB}  are linear in $\bm p\equiv(A,\bm\alpha)$ and the prior on the angles is Gaussian, so $-2\ln\mathcal P$ is quadratic in $\bm p$.
The fitted parameters $\hat{\bm p}$ are then the solution of a linear system, and the posterior of $\bm p$ is Gaussian with covariance denoted by $\Sigma$ (see Appendix~\ref{app:posterior}).
We solve the system iteratively, repeating until the inferred angles no longer change~\cite{MinamiKomatsu2020a,DPK2025} in successive iterations.
For each mass we obtain $\hat{\bm p}=(\hat A,\hat{\bm \alpha})$, $\Sigma$, and  minimum $\chi^2_{\rm min}$ for a given mass, the first term of Eq.~\eqref{eq:posterior}.
We measure the fit relative to a constant rotation,
\begin{equation}
\Delta\chi^2(\mphi)\equiv\chi^2_{\rm min}(\mphi)-\chi^2_{\rm min}(\log_{10}\mphi^{\rm ref}=-30).
\label{eq:dchi2}
\end{equation}

\section{\label{sec:validation}Validation}

For a sufficiently light-mass field, the analysis must reproduce the constant-angle measurement of Ref.~\cite{DPK2025}, and we use this limit to validate our templates and the modified analysis pipeline.  The field is frozen as long as $H\gg\mphi$. At recombination $H(z_*)\simeq3.4\times10^{-29}$\,eV, so for $\mphi=10^{-30}$\,eV the field starts to roll only at redshift $z\sim50$, long after the CMB is emitted.
All photons from the last-scattering surface are therefore rotated by the same angle, $\beta(z_*)\simeq\bini$, and the templates reduce to those of a uniform rotation. At linear order, the templates in Eq.~(\ref{eq:vEB}) and (\ref{eq:vTB}) reduces to $A\,\mathcal{T}^{EB}_\ell\simeq2\beta(D^{EE}_\ell-D^{BB}_\ell)$ and $A\,\mathcal{T}^{TB}_\ell\simeq2\beta D^{TE}_\ell$.
The corrections are negligible for following reasons:
(i) the rotation at reionization differs from $\beta(z_*)$, but it affects only $\ell\lesssim20$ \cite{Sherwin:2021}, far below our multipole range;
(ii) a uniform rotation commutes with lensing, so rotating before lensing, as in our \texttt{CAMBER} code, is equivalent to rotating the lensed spectra.
For $\log_{10}(\mphi/{\rm eV})=-30$, we find $\brec=0.9988$, so $\hat A$ must equal the constant angle of Ref.~\cite{DPK2025} obtained with the same miscalibration prior.
The prior matters because a uniform rotation of the CMB and the common part of the $\alpha_i$ have the same effect on the spectra, and in the absence of foreground $EB$ and $TB$ the two are separated only by the prior on the $\alpha_i$~\cite{DPK2025}.
We therefore compare with their result for $\Pi_{\rm ind}$, the prior used here. Unless otherwise stated, we use this prior throughout the analysis. 

With the $\log_{10}(\mphi/{\rm eV})=-30$ template, we obtain $A=0.204^\circ\pm0.058^\circ$ and $\alpha_i=0.023^\circ$, $-0.063^\circ$, $0.080^\circ$, $-0.038^\circ$, and $-0.008^\circ$ for the array-bands in the order of Table~\ref{tab:bins}, with $\chi^2=2030.85$ for 1984 degrees of freedom and probability to exceed (PTE) $=22.7\%$.
These values agree with the $\Pi_{\rm ind}$ result of Ref.~\cite{DPK2025} (Table~II therein). The rotation at recombination is $\beta(z_*)=\brec\hat A=0.203^\circ$ which is within $\sim 0.02\sigma$ of the fitted value of rotation as expected for a low-mass field.

This fit also serves as the reference for $\Delta\chi^2$ in Sec.~\ref{sec:profile} since all masses $\log_{10}(\mphi/{\rm eV})\lesssim-28.5$ give the same fit.  $\Delta\chi^2$ measures how much worse a heavier field fits than a constant rotation.

\begin{figure}
\centering
\includegraphics[width=\linewidth]{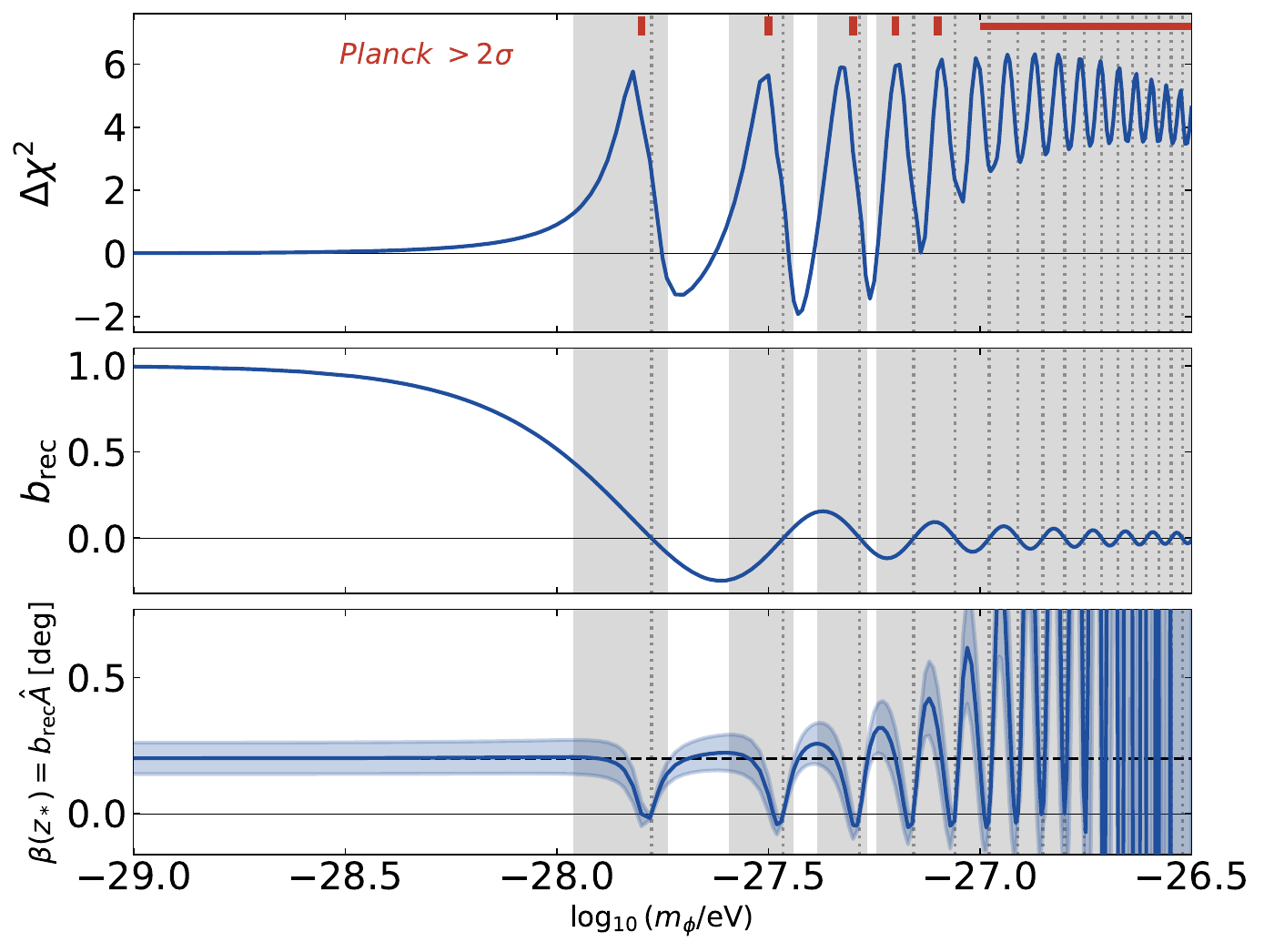}
\caption{\label{fig:scan} Results of the mass scan fitting the ACT DR6 data.
Top: Minimum $\chi^2$ relative to a constant rotation (Eq~\eqref{eq:dchi2}).
Middle: Fraction of the initial rotation that remains at recombination, $\brec=\beta(z_*)/\bini$.
Bottom: Fitted rotation at recombination, $\brec\hat A$, with its $1\sigma$ band.
The dashed horizontal line is the constant-angle result.
In all panels, the gray bands are the masses excluded by the posterior (Eq.~\eqref{eq:excl_bayes}), and the dotted lines mark the zeros-crossing  of $\brec$.
The red marks in the top panel are the masses excluded at more than $2\sigma$ with the \planck\ data~\cite{NMN2025}.
}
\end{figure}

\section{\label{sec:results}Results}
\subsection{\label{sec:profile}Goodness of fit as a function of mass}

Figure.~\ref{fig:scan} shows, as functions of the mass, the minimum $\chi^2$ relative to a constant rotation (top panel), ratio of the rotation at recombination relative to its initial value $b_{\rm rec}$ (middle panel), and the fitted rotation at recombination $\beta(z_*)$  (bottom panel).
The results fall into three mass regimes.

\textit{Light masses}: For $\log_{10}(\mphi/{\rm eV})\lesssim-28.5$,
the field is frozen during recombination ($\brec>0.94$), the templates have the shape of a constant rotation, i.e., $\mathcal{T}^{EB}_\ell\propto2(C^{EE}_\ell-C^{BB}_\ell)$. The fit is as good as for a constant birefringence angle with $\Delta\chi^2<0.05$. All of them give $\beta(z_*)\simeq0.2^\circ$. Therefore ACT DR6 data cannot distinguish these masses.

\textit{Intermediate masses}: For $-28.5\lesssim\log_{10}(\mphi/{\rm eV})\lesssim-27.8$,
the field starts to roll during recombination and $\brec$ falls to zero at $-27.78$.
The fit compensates with a larger amplitude, e.g., $\hat A=0.40^\circ\pm0.12^\circ$ at $-28.0$, which keeps $\beta(z_*)\simeq0.2^\circ$, but the shape of the templates no longer matches the data, and $\Delta\chi^2$ rises to $5.8$ at $-27.82$. 

\textit{Heavy masses}: For $\log_{10}(\mphi/{\rm eV})>-27.78$ the field has crossed zero at least once before recombination, and for $\log_{10}(\mphi/{\rm eV})\gtrsim-27.0$ the field has oscillated several times before recombination. The heavier the field, the more oscillations it completes before
recombination. Therefore $\brec$ passes through zero again and again as the mass increases. $\Delta\chi^2$ follows the same pattern that oscillates in step with $\brec$. It has a total of 15 maxima of values $\sim 5.2$--$6.3$, at
\begin{equation}
\begin{split}
\log_{10}(\mphi/{\rm eV})\simeq{}&-27.82,\;-27.50,\;-27.33,\\
&-27.19,\;-27.09,
\end{split}
\label{eq:peaks}
\end{equation}
and at 11 further masses, separated by $0.03$--$0.08$, in $[-27.0,-26.5]$.
Since the field oscillates during recombination for these masses, the CMB photons last scattered at
slightly different times during recombination are then rotated by angles of
opposite sign. Their contributions to $EB$ and $TB$ partially cancel, and no
amplitude $\hat A$ reproduces the observed spectra. In Figure.~\ref{fig:scan} each
maximum of $\Delta\chi^2$ lies just before a zero crossing of $\brec$ (vertical dotted
lines), that is, where the field is about to change sign at $z_*$.
Every mass that is not excluded by the Bayesian analysis of Sec.~\ref{sec:bayes},
i.e., outside the gray bands in Figure.~\ref{fig:scan}, fits the data equally well as a constant
rotation. $\Delta\chi^2\le1.3$ throughout for those masses, with minima of $-1.31$, $-1.93$ and $-1.44$
at $\log_{10}(\mphi/{\rm eV})=-27.70$, $-27.43$ and $-27.26$. For all these masses the
rotation at recombination is $\beta(z_*)=0.15^\circ$--$0.32^\circ$
which are within  $\sim1\sigma$ of the constant-angle value
$0.204^\circ\pm0.058^\circ$. Above $\log_{10}(\mphi/{\rm eV})=-27.0$ the fit is worse everywhere
($\Delta\chi^2>2.6$), and the whole range up to the end of our scan, $-26.5$, is
excluded at $95\%$ confidence level (Sec.~\ref{sec:bayes}).  

\begin{figure}[h]
\centering
\includegraphics[width=\linewidth]{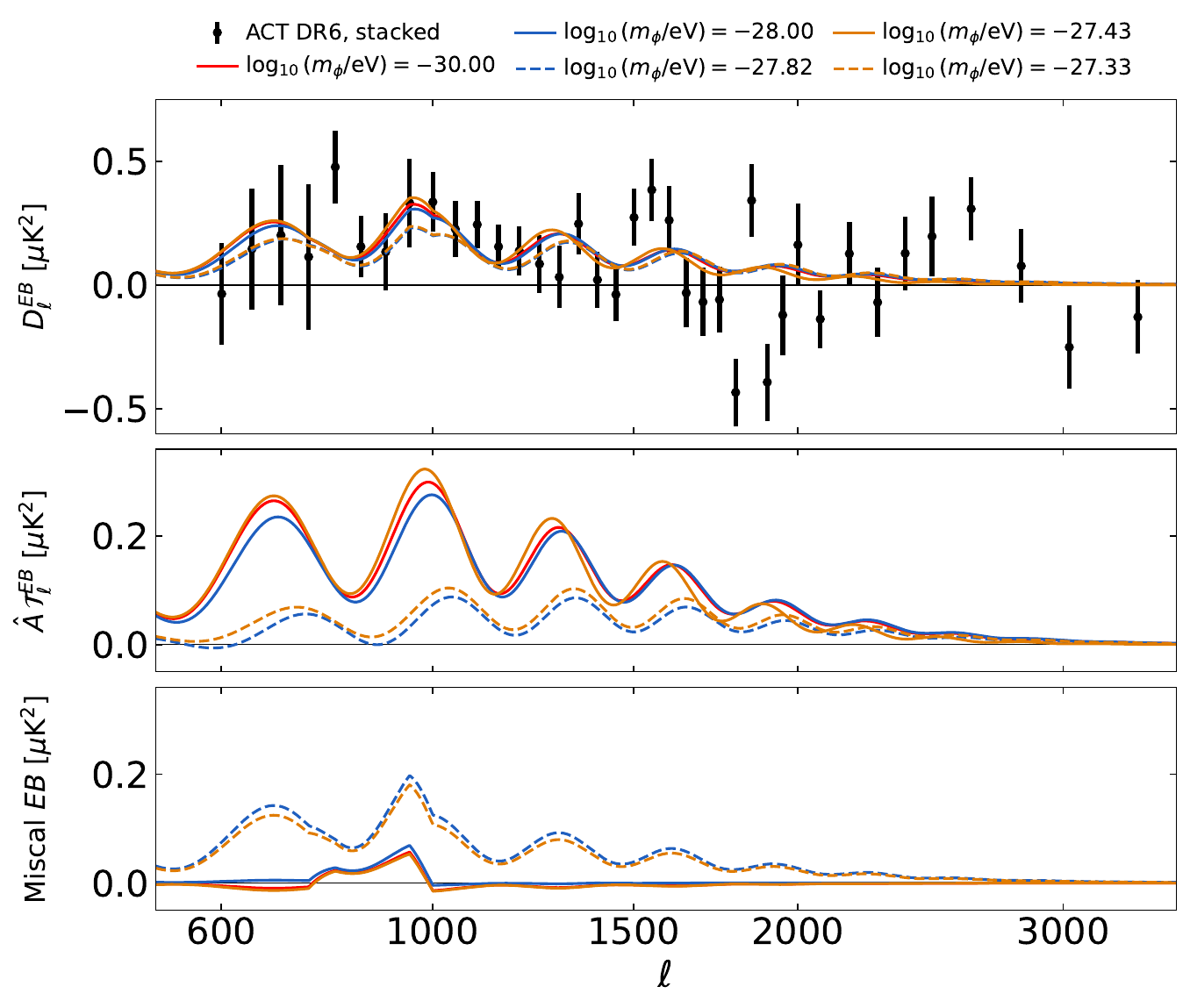}
\caption{\label{fig:stacked} 
ACT DR6 $EB$ spectrum compared with the best-fit models along with its two parts for five axion masses: one light mass (red, $\log_{10}(\mphi/{\rm eV})=-30.00$), two intermediate masses (blue, $-28.00$ and $-27.82$) and two heavy masses (orange, $-27.43$ and $-27.33$). Solid lines are masses that fit as well as the constant-angle model ($\Delta\chi^2=0$, $+0.4$ and $\simeq0$ for $-30.00$, $-28.00$ and $-27.43$). Dashed lines are two local maxima of $\Delta\chi^2$ ($+5.8$ at
$-27.82$ and $+5.2$ at $-27.33$). Top: stacked data and best-fit models. Middle: birefringence part of the model, $\hat A\,\mathcal{T}^{EB}_\ell$. Bottom: miscalibration part of the model, i.e.\ the $EB$ signal from the best-fit miscalibration angles $\alpha_i$. The model in the top panel is the sum of the middle and bottom panels.
}
\end{figure}

To investigate why some masses fit the data poorly, Figure.~\ref{fig:stacked} compares the
stacked ACT DR6 $EB$ spectrum with the best-fit models for five masses. The full
$EB$ and $TB$ comparison is given in Figure.~\ref{fig:stacked_EB_TB}
of Appendix~\ref{app:stacked}.  For the light-mass regime we choose one mass (solid black), and for the intermediate- and heavy-mass
regimes two masses each: one that fits equally well as the constant angle (solid blue and orange) and one
at a local maximum of $\Delta\chi^2$ (dashed blue and orange).
In the top panel, all five best-fit total models agree within the error bars, so the stacked
$EB$ spectrum alone cannot tell them apart. The difference lies in how each model
is split into its two parts.
For the good fits (solid lines), the birefringence part (middle panel) carries the signal. Its
amplitude and acoustic structure are similar the constant-angle template, $D_{\ell}^{EE} -D_{\ell}^{BB}$, and the
miscalibration part (bottom panel) is small.
For masses (dashed lines) corresponding to the local maxima of $\Delta\chi^2$ the template no longer has similar acoustic structure. At $\log_{10}(\mphi/{\rm eV})=-27.82$ and $-27.33$ the birefringence part of the model (middle panel) reaches only $24\%$ and $32\%$ respectively of the birefringence part for the fit at $\log_{10}(\mphi/{\rm eV})=-30$ (red).  The fit makes up the missing $EB$ power with the miscalibration angles. For instance, the miscalibration $EB$ reaches
$\simeq0.2\,\mu{\rm K}^2$ near $\ell\simeq800$--$1000$ for $\mphi/{\rm eV} =-27.82, -27.33$. As a result, in the stacked spectra the total model barely changes  as seen in the Figure.~\ref{fig:stacked} and Figure.~\ref{fig:stacked_EB_TB}. The changes in fitted $\alpha_i$ are spread over many array pairs, some of which fit better and some worse (see Table~\ref{tab:stacked}). This is expected, because each $\alpha_i$ rotates only its own array, whereas birefringence rotates all arrays equally. this causes its changes in $\chi^2$ from $1.47$ at
$-30$ to $5.40$ and $4.50$ at $-27.82$ and $-27.33$ as quoted in Table~\ref{tab:stacked}.


\subsection{\label{sec:bayes}Posterior of the mass}

$\Delta\chi^2$ measures the quality of the best fit but not how much the amplitude must be tuned to achieve it.
To constrain the mass, accounting for the tuning of the amplitude, and compare directly with Ref.~\cite{NMN2025}, we compute the marginal posterior density of the mass with their prescription.
Denoting $\mu\equiv\log_{10}(\mphi/{\rm eV})$,
\begin{equation}
P(\mu\,|\,d)\propto\pi(\mu)\int d^6p\;\mathcal L(d\,|\,\bm p,\mu)\,\pi_\alpha(\bm\alpha)\,\pi(A\,|\,\mu),
\label{eq:bayes}
\end{equation}
where $-2\ln\mathcal L$ is the sum of the first two terms of Eq.~\eqref{eq:posterior}, $\pi_\alpha\propto\exp(-\bm\alpha^{\sf T}\bm\Pi^{-1}\bm\alpha/2)$, and $\pi(\mu)$ is flat in $[-30,-26.5]$.
For the amplitude, $A = \bini$, we adopt the prior of Ref.~\cite{NMN2025}, flat in the rescaled amplitude parameter of
the $EB$ signal,
\begin{equation}
\begin{split}
A_{\rm EB}&\equiv\frac{\bini}{0.2^\circ}F_{\rm sup}(\mphi),\\
F_{\rm sup}&=\left\langle\frac{|\mathcal T^{EB}_\ell(\mphi)|}{|\mathcal T^{EB}_\ell(\mphi^{\rm ref})|}\right\rangle_{600\le\ell\le7525},
\end{split}
\label{eq:aeb}
\end{equation}
which approximates the Jeffreys prior on $\bini$ with $\log_{10}(\mphi^{\rm ref}/{\rm eV})=-30$ (see Appendix.C of~\cite{NMN2025}). Because we integrate over
$A=\bini$ analytically rather than sampling $A_{\rm EB}$, we write this prior in terms
of $A$. At fixed mass, $A_{\rm EB}\propto A$, so
\begin{equation}\label{eq:A_EB_prior}
\pi(A\,|\,\mu)=\pi(A_{\rm EB})\left|\frac{dA_{\rm EB}}{dA}\right|\propto F_{\rm sup}(\mphi).
\end{equation}
In this new-definition, $\pi(A\,|\,\mu)$ is independent of $A$ but depends on the mass through $F_{\rm sup}$. This is
equivalent to sampling $A_{\rm EB}$ with a flat prior, and differs from a flat prior
on $\bini$, for which the factor $F_{\rm sup}$ would be absent. We take the reference angle to be $0.2^\circ$ obtained in Ref.~\cite{DPK2025}, instead of their $0.3^\circ$ obtained from \planck\ data \cite{Eskilt2026}, so that $A_{\rm EB}\simeq1$ for light masses. Although our fit uses both $EB$ and $TB$ that multiply the same amplitude $A=\bini$ parameter with respective templates, we define only $A_{\rm EB}$, and hence the prior,
from the $EB$ template alone, exactly as in Ref.~\cite{NMN2025}, so that the two analyses use the same prior.

\begin{figure}
\centering
\includegraphics[width=\linewidth]{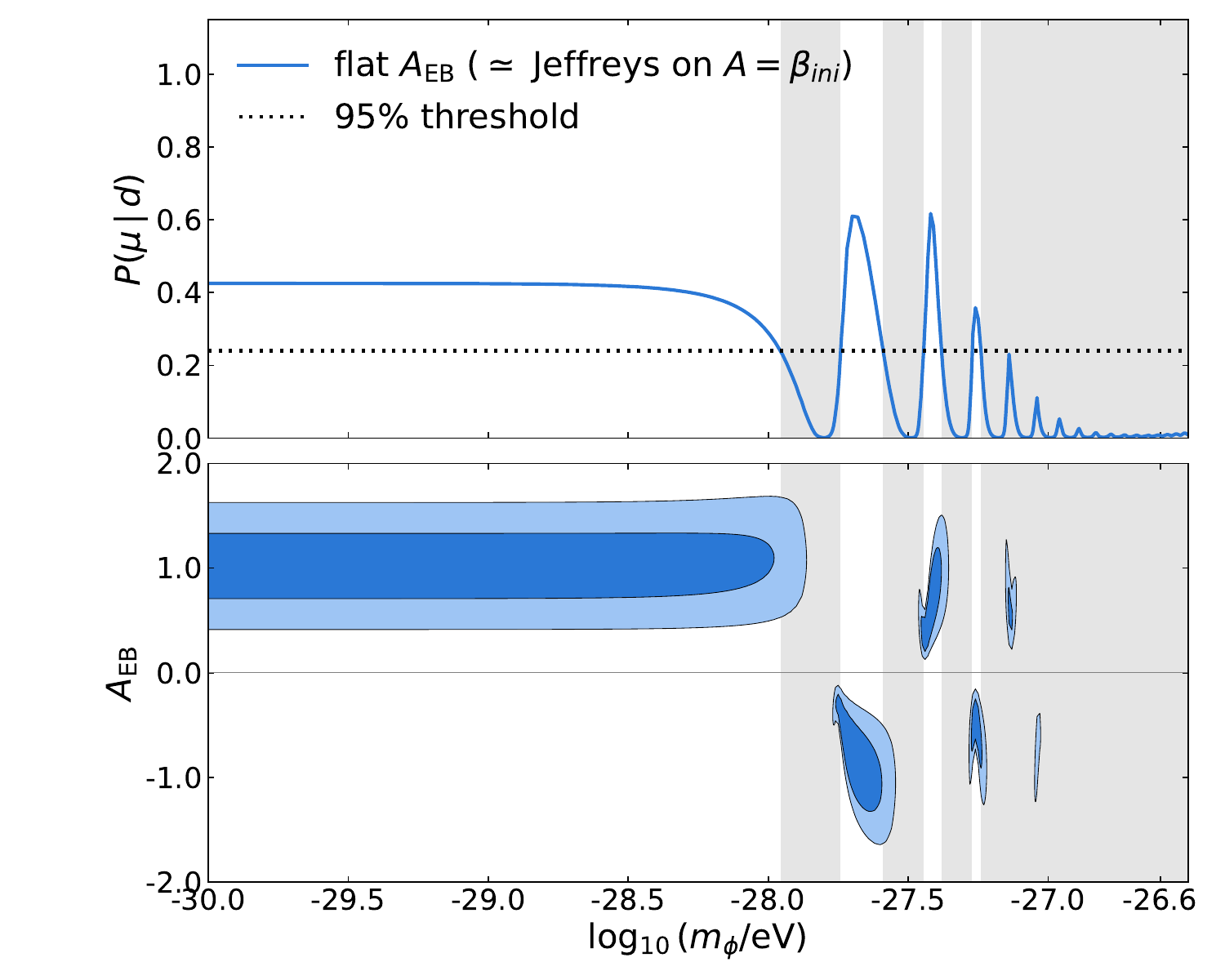}
\caption{\label{fig:bayes} Top: Marginal posterior density of the mass in Eq.~\eqref{eq:post_mu}, for a flat prior on $A_{\rm EB}$. The dotted line is the $95\%$ threshold. Bottom: Joint 68\% and 95\% regions of the mass and $A_{\rm EB}$. The shaded intervals are the masses excluded by the one-dimensional posterior with 95\% confidence. }
\end{figure}

Since the residuals in Eqs~\eqref{eq:vEB} and ~\eqref{eq:vTB} are linear in $\bm p = (A, {\bm\alpha})$ and $\pi_\alpha$ is Gaussian, the integral in Eq.~\eqref{eq:bayes} is analytic (see Appendix~\ref{app:posterior} for detail derivation):
\begin{equation}
\begin{split}
\ln P(\mu\,|\,d)={}&-\tfrac12\left[\chi^2_{\rm min}+\chi^2_{\alpha,{\rm min}}+\ln|\mathbf M|\right]\\
&+\tfrac12\ln\det\Sigma+\ln F_{\rm sup}(\mphi)+{\rm const},
\end{split}
\label{eq:post_mu}
\end{equation}
where $\chi^2_{\alpha,{\rm min}}=\hat{\bm\alpha}^{\sf T}\bm\Pi^{-1}\hat{\bm\alpha}$ with best-fit angles $\hat{\bm\alpha}$.
The terms in brackets measure the quality of the best fit to the data, including the prior on the angles.
The term $\tfrac12\ln\det\Sigma$ measures the volume of parameters the data allow around the best fit, and $\ln F_{\rm sup}$ penalizes masses that require a large initial rotation angle.
We normalize $P(\mu\,|\,d)$ and exclude the masses outside the $95\%$ highest-posterior-density region, $P(\mu\,|\,d)<P_{95}$, where
\begin{equation}
\int_{P(\mu|d)\ge P_{95}}P(\mu\,|\,d)\,d\mu=0.95.
\label{eq:hpd}
\end{equation}
At fixed mass the posterior of $A$ is Gaussian, so the joint posterior is
\begin{equation}
\begin{split}
P(\mu,A_{\rm EB}\,|\,d)&=P(\mu\,|\,d)\;\mathcal N\!\left(A_{\rm EB};\,c\hat A,\,c\,\sigma_A\right),\\
c&\equiv F_{\rm sup}/0.2^\circ.
\end{split}
\label{eq:joint_post}
\end{equation}
Figure.~\ref{fig:bayes} shows the result. The top panel presents the posterior density of the mass, with the excluded masses shown by shaded areas. The bottom panel presents the joint posterior of the mass and the amplitude $A_{\rm EB}$.
For $\mu\lesssim-28.3$ the posterior density is flat, with $A_{\rm EB}=1.02\pm0.29$. These masses all fit the data equally well as constant rotation, so the data cannot distinguish between them and the posterior is flat, like the prior.
With $P_{95}=0.240$, the masses excluded at more than 95\% confidence are
\begin{equation}
\begin{split}
\log_{10}(\mphi/{\rm eV})\in{}&[-27.96,-27.74],\;[-27.59,-27.45],\\
&[-27.38,-27.27],\;[-27.24,-26.5].
\end{split}
\label{eq:excl_bayes}
\end{equation}
Most excluded masses are those that fit the data poorly. The first three excluded
intervals contain the peaks of $\Delta\chi^2$ in Figure.~\ref{fig:scan}. For example, at $\mu=-27.82$, $\Delta\chi^2=5.8$ and the posterior density drops to $P=0.005$. 
 
A few excluded masses fit the data as well as a constant rotation, but need an unusually large initial angle.
For instance, at $\mu=-27.14$, the fit gives $\Delta\chi^2=0.02$, but the field is almost zero at recombination, so an initial angle of $6.2^\circ \pm 1.79^\circ$ is needed to produce the observed $EB$ and $TB$. The prior on $A_{\rm EB}$ in Eq.~\eqref{eq:aeb} disfavors such large initial angles, and the posterior density falls just below the threshold ($P=0.231<P_{95}=0.240$).

Three narrow ranges of heavy mass remain allowed, centered around $\mu=-27.70$, $-27.42$ and $-27.26$, holding $8\%$, $3.5\%$ and $1.5\%$ of the probability. Each lies just above a zero of $\brec$ (see Figure.~\ref{fig:scan}), that is, at masses for which the field has passed through zero shortly before recombination. The sign of $A_{\rm EB}$ alternates between them, $-0.78\pm0.21$, $+0.65\pm0.17$ and $-0.44\pm0.12$, because $\brec$ changes sign from one range to the next.

\begin{figure}
\centering
\includegraphics[width=\linewidth]{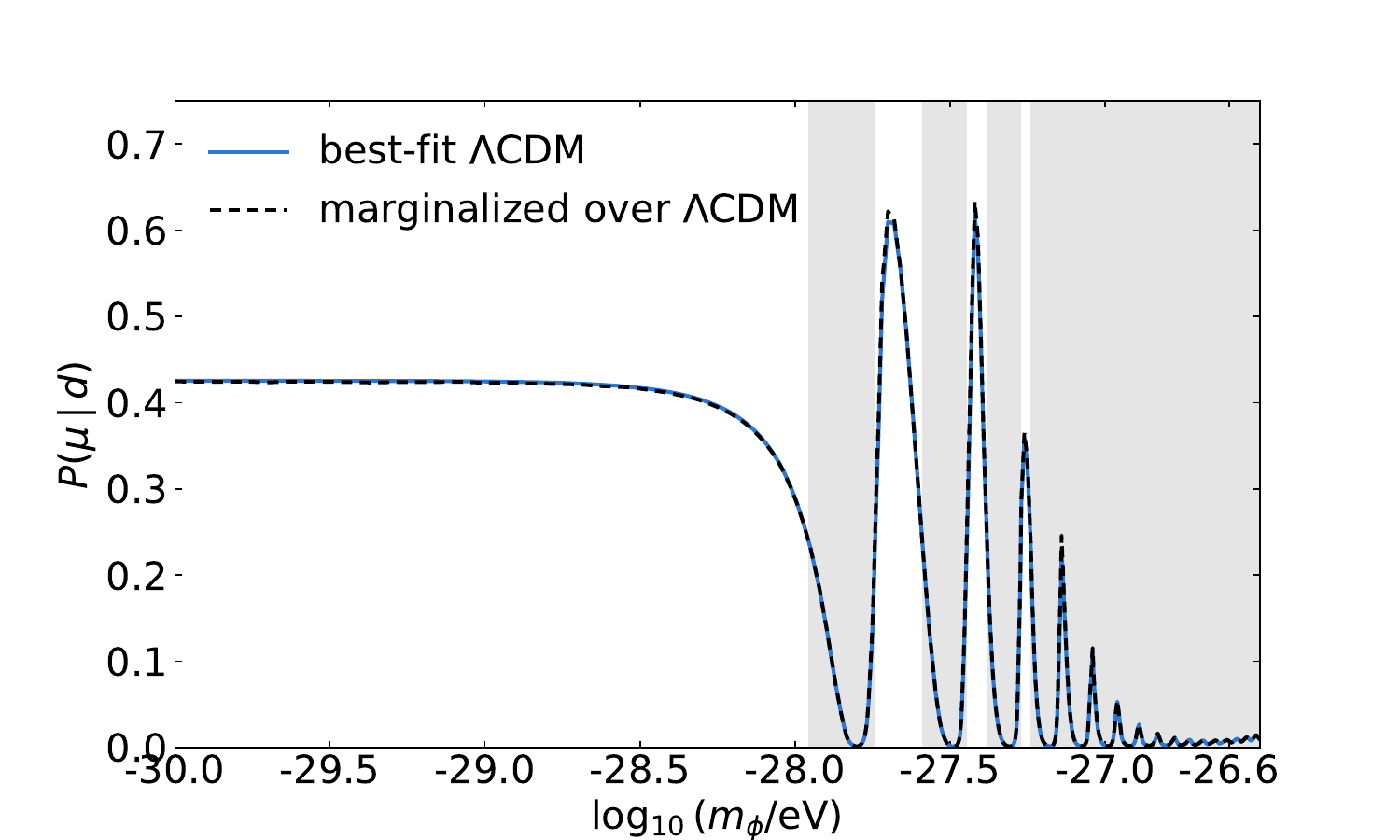}
\caption{\label{fig:lcdm_mass} Marginal posterior of the mass at the ACT DR6 best-fit cosmology (solid), as in Figure.~\ref{fig:bayes}(Top panel), and marginalized over the $\Lambda$CDM parameters with 10,000 samples of the ACT DR6 chain (dashed). The shaded intervals are the masses excluded at the best-fit cosmology.}
\end{figure}
\subsection{\label{sec:cosmology}Dependency on the cosmological parameters}
To test whether fixing the cosmology to its best-fit values affects our results, we marginalize over the $\Lambda$CDM parameters. The detailed calculation of the posterior is given in Appendix.~\ref{app:lcdm}. We draw 10,000 samples of the $\Lambda$CDM parameters from the ACT DR6 $\Lambda$CDM parameter chains~\cite{ACTDR6Louis} and pair each with a mass drawn uniformly from $\mu\in [-30,-26.5]$. For each pair, we compute the templates, fit the data, and weight the pair following Eq.~\eqref{eq:lcdm_weight}. The marginalized mass posterior is nearly identical to the one at the best-fit cosmology as shown in Figure.~\ref{fig:lcdm_mass}. The probabilities of the allowed and excluded intervals change by less than $0.3$\%, and the excluded intervals move by less than $0.002$. Figure.~\ref{fig:lcdm} in Appendix.~\ref{app:lcdm} shows the 68\% and 95\% regions of the $\Lambda$CDM parameters, $\mu$, and $A_{\rm EB}$. The correlation coefficients between the $\Lambda$CDM parameters and the axion parameters do not exceed $0.044$ in absolute value. Within the ACT uncertainties, the cosmology therefore does not affect the constraints on the axion parameters.


\subsection{\label{sec:planck}Comparison with \planck}

Ref.~\cite{NMN2025} analyzed the \planck\ $EB$ power spectra at $51\le\ell\le1490$ with a model of the dust $EB$ correlation and flat priors on the miscalibration angles.
They excluded $\log_{10}(\mphi/{\rm eV})\simeq-27.8$, $-27.5$, $-27.3$, $-27.2$, and $-27.1$ and the interval $[-27.0,-26.5]$ at more than $2\sigma$, and found masses below about $10^{-28}$\,eV unconstrained. These mass ranges are marked by red in the top panel of Figure.~\ref{fig:scan}.
Our maxima of $\Delta\chi^2$ coincide with these masses as shown in Figure.~\ref{fig:scan}, and all of them lie within our excluded intervals.
The two analyses differ in the experiment, the multipole range, the spectra fitted ($EB$ alone in Ref.~\cite{NMN2025} vs. $EB$ and $TB$ in this work), and the way the degeneracy with the miscalibration angles is broken (dust intrinsic $EB$ in Ref.~\cite{NMN2025} vs. instrument model in this work). So their agreement in results indicates that the pattern is a robust consequence of the field dynamics.
In the constant-angle analyses for low-mass fields, the inferred rotation angles differ in two analyses: $\simeq0.36^\circ$ for \planck\ and $\simeq0.20^\circ$ for ACT. The same difference is also reported in Ref.~\cite{Eskilt2026}

\section{\label{sec:discussion}Discussion and conclusions}
We have constrained the mass of an axion-like field that could cause the isotropic cosmic birefringence found in the ACT DR6 polarization data.
If the field starts to oscillate before or during recombination, the rotation angle changes while the CMB is emitted, and the $EB$ and $TB$ spectra are no longer proportional to the $EE-BB$ and $TE$ spectra, which are the standard templates for a constant angle rotation.
We computed the $EB$ and $TB$ templates with a modified version of \texttt{CAMB} for masses $10^{-30}$--$10^{-26.5}$\,eV and a quadratic potential, and fitted them to the ACT DR6 $EB$ and $TB$ spectra, together with the miscalibration angles of each array-band, using the Minami--Komatsu estimator~\cite{MinamiKomatsu2020a} as implemented in Ref.~\cite{DPK2025}.
Our main results are as follows.
\begin{itemize}
\item For $\log_{10}(\mphi/{\rm eV})\lesssim-28.5$, the field is still frozen at recombination, and every mass produces the same constant rotation $\beta(z_*)\simeq0.2^\circ$ at recombination. These masses are equally likely and contain $84\%$ of the posterior probability.
\item At larger masses, the data allow only three narrow windows, near $\log_{10}(\mphi/{\rm eV})\simeq-27.70$, $-27.42$, and $-27.26$.
\item  Masses in the intervals $\log_{10}(\mphi/{\rm eV})\in$ $[-27.96,-27.74]$, $[-27.59,-27.45]$, $[-27.38,-27.27]$, and $[-27.24,-26.5]$
are excluded at more than 95\% confidence. The first three contain the masses for which the field crosses zero during recombination, where the signal changes shape, and the fit worsens by $\Delta\chi^2\simeq5$--$6$ and in the last interval, the data can be fitted only with a large initial angle.
\end{itemize}
These intervals change by at most $0.002$ in $\log_{10}\mphi$ when marginalized over the $\Lambda$CDM parameters. Therefore fixing the cosmology to its best-fit values does not affect our conclusions. The masses that are excluded with $2\sigma$ analyzing the \planck\ data are $\log_{10}(\mphi/{\rm eV})\simeq-27.8$, $-27.5$, $-27.3$, $-27.2$, $-27.1$, and $[-27.0,-26.5]$~\cite{NMN2025}. All lie within our excluded intervals, and both analyses leave the lighter masses unconstrained. These constraints assume a  flat prior in rescaled amplitude $A_{\rm EB}$, which approximates the Jeffreys prior on our fitted amplitude parameter $A=\bini$.  A flat prior in $\bini$ would instead favor heavy masses simply because their templates are small and leave $\bini$ unconstrained. 
We used independent priors on the miscalibration angles of Ref.~\cite{DPK2025}. Other priors, e.g., $\Pi_{\rm PA}$ of Ref.~\cite{DPK2025} which correlates the angles within each array, may change our constraints. As in Ref.~\cite{DPK2025}, we also neglect the foreground $EB$ and $TB$ correlations at the ACT angular scales and do not marginalize over the remaining $I\to P$ leakage. An analysis with the $\Pi_{\rm PA}$ prior and a marginalization over the intrinsic $EB$ and $TB$ correlations of the dust are left for future work.

Three extensions would sharpen these results. A joint analysis of ACT and \planck\ would combine their complementary ranges of multipoles.
Extensions of $\Lambda$CDM that change the recombination history, such as early dark energy or a varying electron mass, change the shape of the signal and we leave these for future work. Finally, all masses considered here predict no rotation of the reionization $EB$ signal, whereas a lighter field rotates it by the same $\simeq0.2^\circ$ as at recombination. Since a miscalibration rotates all multipoles equally, a measurement of $EB$ at $\ell\lesssim30$, e.g., by LiteBIRD, would test this independently of the calibration of the instruments.

\begin{acknowledgments}
DA acknowledges financial support from the projects PID2023-150398NBI00 and PID2023-151567NB-I00, funded by the Ministry of Science, Innovation and Universities of Spain, AEI/10.13039/501100011033 and by FEDER, EU. DA thanks P.~Diego-Palazuelos and E.~Komatsu for making their birefringence code public on which this analysis is based, and the ACT Collaboration for the public DR6 data products. Some of the results in this paper have been derived using the  \texttt{NumPy} \cite{Numpy:2020}, \texttt{SciPy} \cite{scipy:2020}, and \texttt{Matplotlib}\cite{Matplotlib:2007} packages.
\end{acknowledgments}

\appendix

\section{\label{app:stacked} Stacked power spectra}

The fit uses the $EB$ and $TB$ spectra of all array-band pairs with the full covariance.
For display only, we combine the pairs into a single spectrum per bin.
In each bin $b$, the band powers $D_{b,k}$ of the pairs $k$ that contain that bin are combined with the minimum-variance weights
\begin{align}
\bar D_b&=\bm w_b^{\sf T}\bm D_b,\qquad
\bm w_b=\frac{\mathbf C_b^{-1}\bm u}{\bm u^{\sf T}\mathbf C_b^{-1}\bm u},\\
\sigma^2(\bar D_b)&=\frac{1}{\bm u^{\sf T}\mathbf C_b^{-1}\bm u},
\end{align}
where $\mathbf C_b$ is the covariance between the pairs in bin $b$; correlations between bins are neglected and $\bm u=(1,\dots,1)^{\sf T}$.

\begin{figure*}  
\centering
\includegraphics[width=\textwidth]{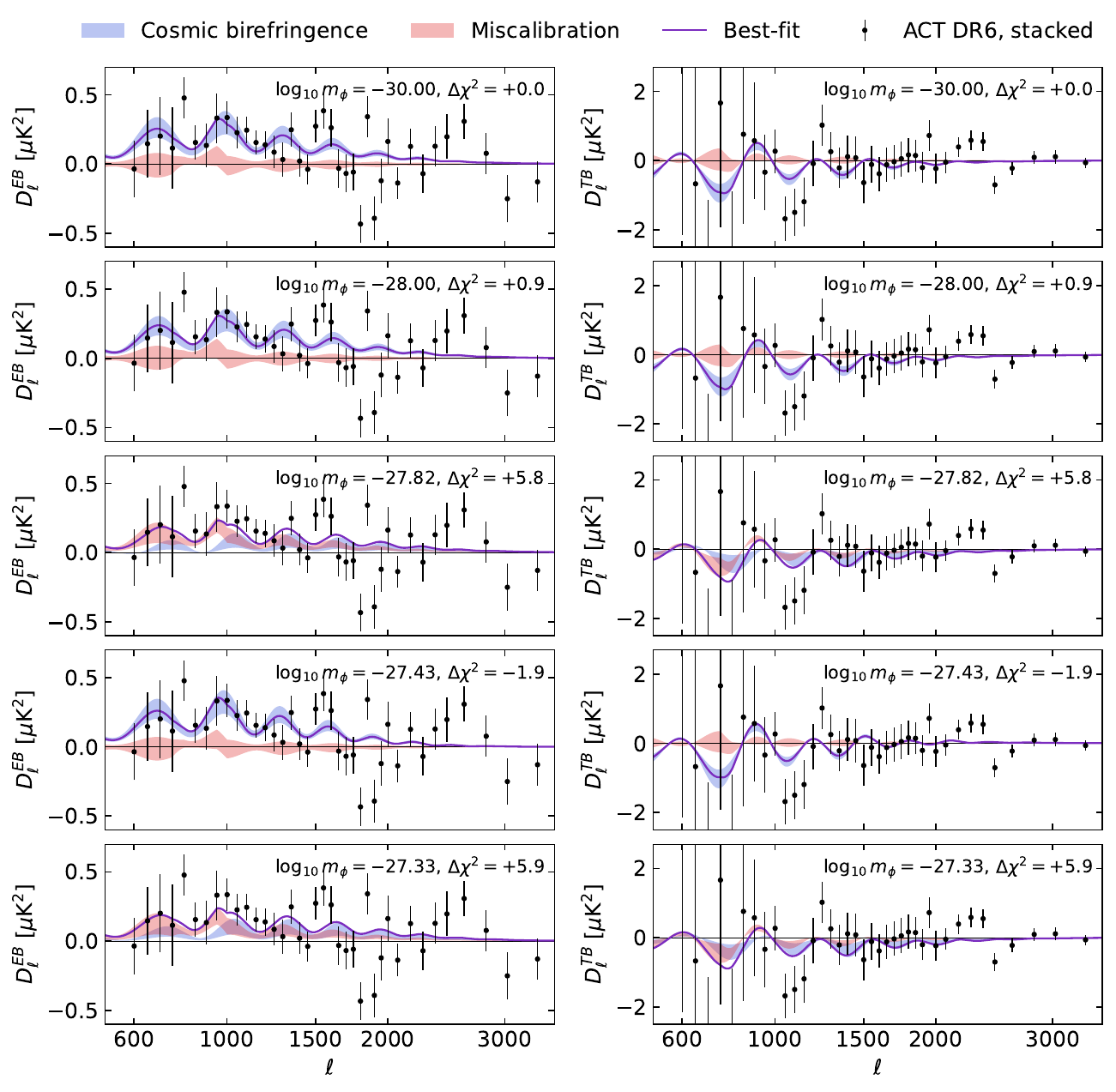}
\caption{Stacked ACT DR6 $EB$ (left) and $TB$ (right) spectra, with the best-fit
model for five axion masses (one mass per row). The blue band is $1\sigma$ spread of the birefringence
part of the model, the red band is $1\sigma$ spread of the miscalibration part,  and the
purple line is the best-fit from both contributions. At $\log_{10}(\mphi/{\rm eV})=-27.82$ and $-27.33$,  $\Delta\chi^2$ is maximum, and the birefringence part is small, and the miscalibration angles produce most of the signal.}
\label{fig:stacked_EB_TB}
\end{figure*}

Figure.~\ref{fig:stacked_EB_TB} shows the stacked $EB$ and $TB$ spectra with the best-fit model for the five masses, and also split into its two parts.
The birefringence part, $\hat A\,\mathcal T^{XB}_\ell(\mphi)$, is the same for every pair.
The miscalibration part is the signal from the best-fit angles $\hat\alpha_i$. 
The bands show the $1\sigma$ spread of each part.

Table~\ref{tab:stacked} lists the best-fit parameters for these five masses. For the masses that fit equally well as a constant rotation ($\mu=-30$, $-28.00$, and $-27.43$), the angles agree within $0.01^\circ$ and $\chi^2_\alpha\simeq1.5$. However at the maxima of $\Delta\chi^2$ ($\mu=-27.82$ and $-27.33$), the birefringence part is too small to reproduce the data, and all five angles shift by $+0.01^\circ$ to $+0.12^\circ$, mimicking a common rotation.
This shift costs $\chi^2_\alpha=5.40$ and $4.50$, and the fit to the spectra also worsens, giving the high $\Delta\chi^2$ shown in Figure.~\ref{fig:scan}.
\begin{table*}
\caption{\label{tab:stacked} Best-fit parameters and $\chi^2$ values for the five masses of Figures.~\ref{fig:stacked} and~\ref{fig:stacked_EB_TB}; $\beta(z_*)$, and the uncertainties are the square roots of the diagonal of $\Sigma$. $\Delta\chi^2$ is relative to $\log_{10}(\mphi/{\rm eV})=-30$, and $\chi^2_\alpha=\hat{\bm\alpha}^{\sf T}\bm\Pi^{-1}\hat{\bm\alpha}$ is the contribution of the prior on the miscalibration angles. At the two maxima of $\Delta\chi^2$ (for masses $-27.82$ and $-27.33$), $\beta(z_*)$ is smaller than the constant-angle value, and the angles of PA5 and PA6 shift together toward positive values to compensate, which raises $\chi^2_\alpha$.}
\begin{ruledtabular}
\begin{tabular}{cccccccccc}
$\log_{10}(\mphi/{\rm eV})$ & $\hat A$ & $\beta(z_*)$ & $\Delta\chi^2$ & $\chi^2_\alpha$ &
$\alpha_{\rm PA4f220}$ & $\alpha_{\rm PA5f090}$ & $\alpha_{\rm PA5f150}$ & $\alpha_{\rm PA6f090}$ & $\alpha_{\rm PA6f150}$ \\
\hline
$-30.00$ & $0.204\pm0.058$ & $0.204\pm0.058$ & $0$     & $1.47$ & $+0.02\pm0.09$ & $-0.06\pm0.06$ & $+0.08\pm0.06$ & $-0.04\pm0.07$ & $-0.01\pm0.07$ \\
$-28.00$ & $0.40\pm0.12$   & $0.21\pm0.06$   & $+0.91$ & $1.53$ & $+0.02\pm0.09$ & $-0.06\pm0.06$ & $+0.09\pm0.06$ & $-0.03\pm0.07$ & $+0.00\pm0.07$ \\
$-27.82$ & $0.94\pm0.57$   & $0.10\pm0.06$   & $+5.77$ & $5.40$ & $+0.04\pm0.09$ & $+0.02\pm0.06$ & $+0.18\pm0.05$ & $+0.05\pm0.07$ & $+0.11\pm0.06$ \\
$-27.43$ & $2.35\pm0.62$   & $0.20\pm0.05$   & $-1.93$ & $1.41$ & $+0.02\pm0.09$ & $-0.06\pm0.06$ & $+0.08\pm0.06$ & $-0.04\pm0.07$ & $-0.01\pm0.07$ \\
$-27.33$ & $1.36\pm0.73$   & $0.16\pm0.08$   & $+5.90$ & $4.50$ & $+0.03\pm0.09$ & $+0.01\pm0.06$ & $+0.17\pm0.06$ & $+0.04\pm0.07$ & $+0.10\pm0.06$ \\
\end{tabular}
\end{ruledtabular}
\end{table*}

\section{\label{app:posterior}Marginal posterior of axion-mass}
We collect the six parameters in $\bm p=(A,\alpha_1,\dots,\alpha_5)^{\sf T}$.
The residuals of Eqs.~\eqref{eq:vEB} and \eqref{eq:vTB} are linear in $\bm p$, and
stacking them over all bins, array pairs, and the two spectra gives
\begin{equation}
\bm v=\bm d-\mathbf X\bm p .
\label{eq:vlin}
\end{equation}
Here $\bm d$ is the vector of observed $EB$ and $TB$ band powers, and $\mathbf X$ is
a matrix containing the templates and the observed $EE$, $BB$, and $TE$ band powers.

Stacking also the blocks $\mathbf M_{bb'}$ into one matrix $\mathbf M$, we write the first and third terms of Eq.~\eqref{eq:posterior} as
\begin{equation}\label{eq:redefine_chi2}
\chi^2(\bm p)\equiv\bm v^{\sf T}\mathbf M^{-1}\bm v,\qquad
\chi^2_\alpha(\bm p)\equiv\bm\alpha^{\sf T}\bm\Pi^{-1}\bm\alpha .
\end{equation}
Therefore Eq.~\eqref{eq:posterior} reads $-2\ln\mathcal P=\chi^2+\ln|\mathbf M|+\chi^2_\alpha$.
The first two terms are the likelihood, $-2\ln\mathcal L=\chi^2+\ln|\mathbf M|$, and the last is the prior on the miscalibration angles, $\pi_\alpha\propto e^{-\chi^2_\alpha/2}$.
With $\pi(A\,|\,\mu)\propto F_{\rm sup}(\mphi)$ from Eq.~\eqref{eq:A_EB_prior}, finally the integrand of Eq.~\eqref{eq:bayes} becomes proportional to
\begin{equation}
e^{-\frac12\left[\chi^2(\bm p)+\chi^2_\alpha(\bm p)+\ln|\mathbf M|\right]}\,F_{\rm sup}(\mphi).
\label{eq:integrand}
\end{equation}
Rewriting $\chi^2_\alpha=\bm p^{\sf T}\tilde{\bm\Pi}^{-1}\bm p$ with $\tilde{\bm\Pi}^{-1}\equiv{\rm diag}(0,\bm\Pi^{-1})$ due to flat prior for $A$, and expanding sum of Eqs~\eqref{eq:redefine_chi2},
\begin{equation}
\begin{split}
\chi^2(\bm p)+\chi^2_\alpha(\bm p)&=(\bm d-\mathbf X\bm p)^{\sf T}\mathbf M^{-1}(\bm d-\mathbf X\bm p)+\bm p^{\sf T}\tilde{\bm\Pi}^{-1}\bm p\\
&=\bm p^{\sf T}\Sigma^{-1}\bm p-2\,\bm p^{\sf T}\bm B+\bm d^{\sf T}\mathbf M^{-1}\bm d,
\end{split}
\end{equation}
with
\begin{equation}
\Sigma^{-1}\equiv\mathbf X^{\sf T}\mathbf M^{-1}\mathbf X+\tilde{\bm\Pi}^{-1},\qquad
\bm B\equiv\mathbf X^{\sf T}\mathbf M^{-1}\bm d .
\label{eq:SigmaB}
\end{equation}
With $\hat{\bm p}\equiv\Sigma\bm B$,
\begin{equation}
(\bm p-\hat{\bm p})^{\sf T}\Sigma^{-1}(\bm p-\hat{\bm p})=\bm p^{\sf T}\Sigma^{-1}\bm p-2\,\bm p^{\sf T}\bm B+\bm B^{\sf T}\Sigma\bm B,
\end{equation}
so that
\begin{equation*}
\begin{split}
&\chi^2(\bm p)+\chi^2_\alpha(\bm p)-(\bm p-\hat{\bm p})^{\sf T}\Sigma^{-1}(\bm p-\hat{\bm p})\\
&\quad=\bm d^{\sf T}\mathbf M^{-1}\bm d-\bm B^{\sf T}\Sigma\bm B\\
&\quad=\bm d^{\sf T}\mathbf M^{-1}\bm d-2\,\hat{\bm p}^{\sf T}\bm B
+\hat{\bm p}^{\sf T}\mathbf X^{\sf T}\mathbf M^{-1}\mathbf X\hat{\bm p}
+\hat{\bm p}^{\sf T}\tilde{\bm\Pi}^{-1}\hat{\bm p}\\
&\quad=(\bm d-\mathbf X\hat{\bm p})^{\sf T}\mathbf M^{-1}(\bm d-\mathbf X\hat{\bm p})
+\hat{\bm p}^{\sf T}\tilde{\bm\Pi}^{-1}\hat{\bm p}\\
&\quad=\chi^2_{\rm min}+\chi^2_{\alpha,{\rm min}},
\end{split}
\end{equation*}
where $\chi^2_{\rm min}\equiv\chi^2(\hat{\bm p})$ and $\chi^2_{\alpha,{\rm min}}\equiv\chi^2_\alpha(\hat{\bm p})=\hat{\bm\alpha}^{\sf T}\bm\Pi^{-1}\hat{\bm\alpha}$.
Therefore
\begin{equation}
\chi^2(\bm p)+\chi^2_\alpha(\bm p)=\chi^2_{\rm min}+\chi^2_{\alpha,{\rm min}}+(\bm p-\hat{\bm p})^{\sf T}\Sigma^{-1}(\bm p-\hat{\bm p}).
\label{eq:square6}
\end{equation}
Since $\Sigma^{-1}$ is symmetric and positive definite, the last term of Eq~\eqref{eq:square6} is nonnegative and vanishes only at $\bm p=\hat{\bm p}$, so $\hat{\bm p}$ is the maximum-likelihood solution of Eq.~\eqref{eq:posterior} at fixed $\mathbf M$.
Using Eq.~\eqref{eq:square6} in Eq.~\eqref{eq:integrand}, and integrating the only factor that depends on $\bm p$, gives
\begin{equation}
\int d^6p\;e^{-\frac12(\bm p-\hat{\bm p})^{\sf T}\Sigma^{-1}(\bm p-\hat{\bm p})}=(2\pi)^3\sqrt{\det\Sigma}.
\label{eq:gauss6}
\end{equation}
Here, as in the fit, $\mathbf M$ is evaluated at $\hat{\bm p}$ and its variation across the width of the posterior is neglected.
Finally, Eq.~\eqref{eq:bayes} becomes
\begin{equation}
P(\mu\,|\,d)\propto e^{-\frac12\left[\chi^2_{\rm min}+\chi^2_{\alpha,{\rm min}}+\ln|\mathbf M|\right]}\sqrt{\det\Sigma}\;F_{\rm sup}(\mphi),
\end{equation}
whose logarithm is Eq.~\eqref{eq:post_mu}. The factor $(2\pi)^3$ and the normalization of $\pi_\alpha$ are absorbed in normalized $P(\mu\,|\,d)$.


\section{Marginalization over the cosmological parameters}\label{app:lcdm}
We denote $TT$, $TE$, and $EE$ band powers by $d_{\rm ACT}$, from which the ACT DR6 $\Lambda$CDM parameters samples are derived \cite{ACTDR6Louis}. We use notation  $d$ for the $EB$ and $TB$ band powers as in the main text in Sec.~\ref{sec:bayes}, and $\bm\theta$ for the $\Lambda$CDM parameters. In the main text, we fixed the cosmology to its best-fit values $\bm\theta_{\rm bf}$. In this section we explain how we can estimate the joint posterior of $\bm\theta$ and $\mu$ for both data sets, $d_{\rm ACT}$ and $d$, and show that it can be sampled by importance-weighting draws from the ACT DR6 $\Lambda$CDM parameter chains.
With $\bm p=(A,\bm\alpha)$,
\begin{equation}
\begin{split}
P(\bm\theta,\mu,\bm p\,|\,d_{\rm ACT},d)\propto{}&\mathcal L(d_{\rm ACT},d\,|\,\bm p,\mu,\bm\theta)\\
&\times\pi(\bm\theta)\,\pi(\mu)\,\pi_\alpha(\bm\alpha)\,\pi(A\,|\,\mu),
\end{split}
\label{eq:lcdm_bayes}
\end{equation}
where
\begin{equation}
\mathcal L(d_{\rm ACT},d\,|\,\bm p,\mu,\bm\theta)=\mathcal L(d_{\rm ACT}\,|\,\bm p,\mu,\bm\theta)\;\mathcal L(d\,|\,d_{\rm ACT},\bm p,\mu,\bm\theta).
\label{eq:lcdm_product}
\end{equation}
Now we make two assumptions: (a) The rotation leaves $TT$ unchanged and changes $TE$ and $EE$ only at second order in $\beta$, so that $\mathcal L(d_{\rm ACT}\,|\,\bm p,\mu,\bm\theta)\simeq\mathcal L_{\rm ACT}(d_{\rm ACT}\,|\,\bm\theta)$.
(b) The noise of the $EB$ and $TB$ band powers is uncorrelated with that of the $TT$, $TE$, and $EE$ band powers, so that $\mathcal L(d\,|\,d_{\rm ACT},\bm p,\mu,\bm\theta)\simeq\mathcal L(d\,|\,\bm p,\mu,\bm\theta)$ which is same likelihood as in Eq.~\eqref{eq:posterior} with templates computed for $\bm\theta$.
The likelihood $\mathcal L(d\,|\,\bm p,\mu,\bm\theta)$ still depends on $\bm\theta$, through the templates.

By Bayes theorem, $\mathcal L_{\rm ACT}(d_{\rm ACT}\,|\,\bm\theta)\,\pi(\bm\theta)\propto P_{\rm ACT}(\bm\theta)$, the posterior of the $\Lambda$CDM parameters using $d_{\rm ACT}$. Integrating over $\bm p$ as in Eq.~\eqref{eq:bayes}, Eq.~\eqref{eq:lcdm_bayes} becomes
\begin{equation}
P(\bm\theta,\mu\,|\,d_{\rm ACT},d)\propto P_{\rm ACT}(\bm\theta)\,\pi(\mu)\,P(\mu\,|\,d,\bm\theta).
\label{eq:post_lcdm}
\end{equation}

Equation~\eqref{eq:post_lcdm} is the product of two parts. The first part, $P_{\rm ACT}(\bm\theta)\,\pi(\mu)$, is easy to sample. We take $\Lambda$CDM parameter samples $\bm\theta_k$ from the ACT DR6 results. For each $\bm\theta_k$ we draw a mass $\mu_k$ uniformly in $[-30,-26.5]$. The second part, $P(\mu\,|\,d,\bm\theta)$, is the posterior of Eq.~\eqref{eq:post_mu}, computed at the cosmology $\bm\theta_k$. We include it by giving each pair $(\bm\theta_k,\mu_k)$ the weight
\begin{equation}
w_k=P(\mu_k\,|\,d,\bm\theta_k).
\label{eq:lcdm_weight}
\end{equation}
The weighted pairs then follow the joint posterior of Eq.~\eqref{eq:post_lcdm}.  We use 10,000 such $(\bm\theta_k,\mu_k)$ pairs to reduce the computation cost.

The cosmological parameters and the mass are correlated only if the fit at a given mass depends on the fiducial cosmology. If $P(\mu\,|\,d,\bm\theta)$ were a function of $\mu$ times a function of $\bm\theta$, Eq.~\eqref{eq:post_lcdm} would also factorize, and the two would be uncorrelated. The analysis of Sec.~\ref{sec:bayes} is the special case in which the cosmology is fixed, $P_{\rm ACT}(\bm\theta)\equiv \delta(\bm\theta-\bm\theta_{\rm bf})$.

Integrating Eq.~\eqref{eq:post_lcdm} over $\bm\theta$, and writing the result relative to the best-fit cosmology, gives
\begin{equation}
\begin{split}
P(\mu\,|\,d_{\rm ACT},d)&\propto P(\mu\,|\,d,\bm\theta_{\rm bf})\,R(\mu),\\
R(\mu)&=\int d\bm\theta\;P_{\rm ACT}(\bm\theta)\,\frac{P(\mu\,|\,d,\bm\theta)}{P(\mu\,|\,d,\bm\theta_{\rm bf})}.
\end{split}
\label{eq:lcdm_R}
\end{equation}
$R(\mu)$ is the average over the cosmologies allowed by ACT. $R(\mu)$ indicates how much the fit at mass $\mu$ improves or degrades relative to the best-fit cosmology. If $R(\mu)$ is constant, marginalizing over the cosmology leaves the mass posterior unchanged.

We compute $P(\mu\,|\,d,\bm\theta_{\rm bf})$ in Sec.~\ref{sec:bayes}, and here we estimate only $R(\mu)$ from the weighted pairs. 
 Equation~\eqref{eq:lcdm_R} is an average over cosmologies at fixed mass. Each cosmology $\bm\theta_k$, however, is evaluated at a single mass $\mu_k$. We therefore estimate $R(\mu)$ from the pairs whose mass $\mu_k$ lies close to $\mu$. For each pair we compute the ratio $r_k=w_k/P(\mu_k\,|\,d,\bm\theta_{\rm bf})$, and take $R(\mu)$ as the average of $r_k$ over the pairs near $\mu$. Rather than a sharp cut in $|\mu_k-\mu|$, we give pairs closer to $\mu$ a larger weight:
\begin{equation}
R(\mu)\simeq\frac{\sum_k e^{-(\mu_k-\mu)^2/2\sigma^2}\,r_k}{\sum_k e^{-(\mu_k-\mu)^2/2\sigma^2}}
\label{eq:lcdm_Rest}
\end{equation}
with $\sigma=0.03$. This width is large enough to include about 100 pairs $(\bm\theta_k,\mu_k)$ around each $\mu$, and small enough not to blur the allowed mass intervals. Any $\sigma$ between $0.01$ and $0.2$ gives the same results. 
Figure.~\ref{fig:lcdm_mass} compares the result with the posterior at the best-fit cosmology. The probabilities of the allowed and excluded intervals change by less than $0.3$\%. Therefore fixing the cosmology to its best-fit values, as in the main text, does not affect our conclusions. 

\begin{figure*}
\centering
\includegraphics[width=0.85\textwidth]{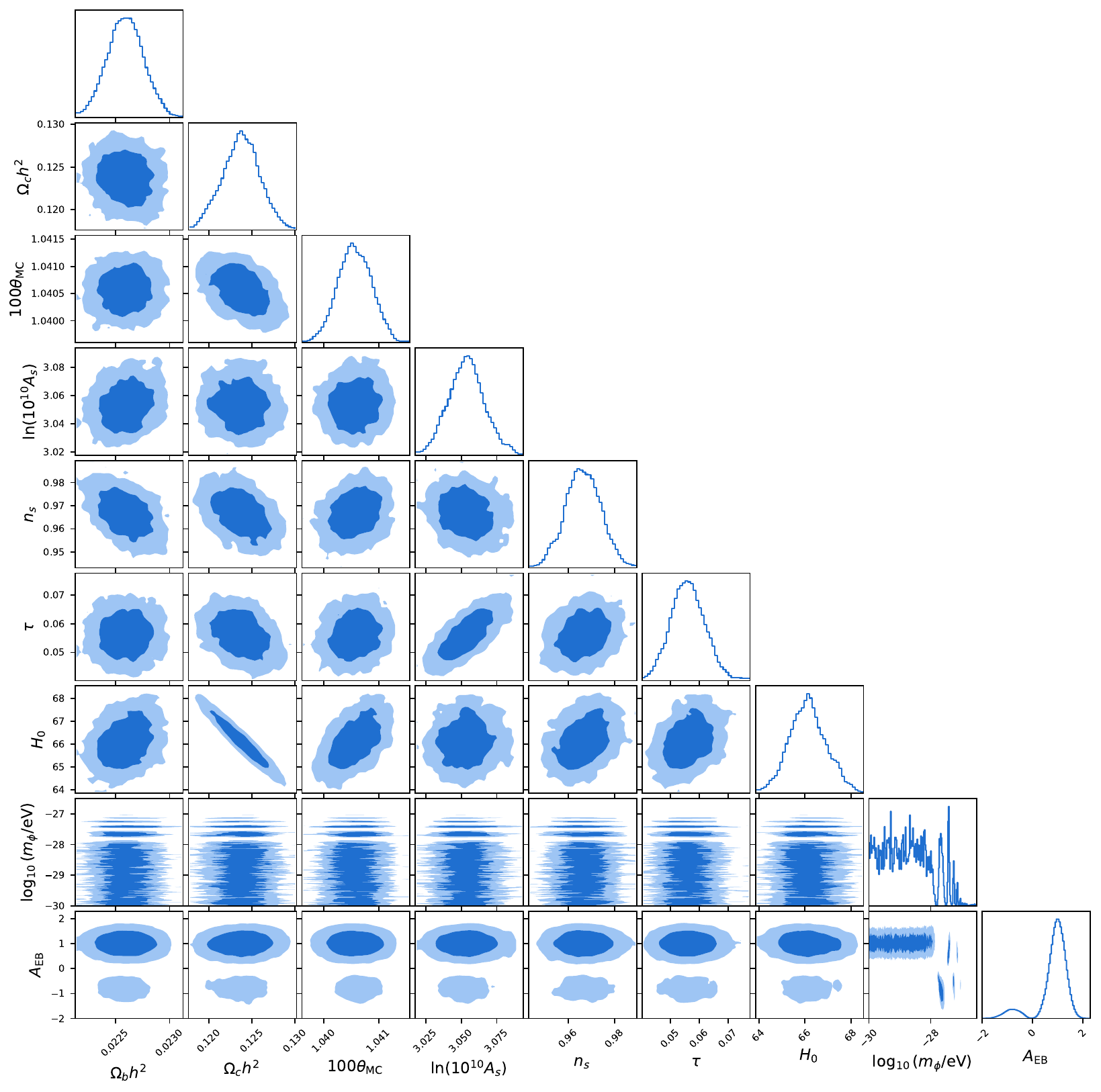}
\caption{\label{fig:lcdm}  68\% and 95\% regions of the joint posterior of the $\Lambda$CDM parameters, the axion mass, and $A_{\rm EB}$. The mass axis is not smoothed, so that the narrow allowed mass intervals remain intact in the figure. The axion parameters show no correlation with the $\Lambda$CDM parameters.} 
\end{figure*}

\bibliographystyle{apsrev4-2}
\bibliography{references}

\end{document}